\documentstyle[12pt]{article}
\begin{document}
\baselineskip=0.89cm
\vspace{4cm}
\begin{center}
{\bf Manifestation of spin degrees of freedom\par
in the double fractional quantum Hall system}\par
\ \\
T. Nakajima$^*$ and H. Aoki\par
{\it Department of Physics, University of Tokyo, Hongo, Tokyo 113, Japan}\par
\end{center}
\vspace{0.4cm}
\par
\noindent {\bf Abstract :}\hspace{4pt}  The double fractional quantum Hall
system of spin 1/2 electrons is numerically studied to predict that there
exists a novel spin-unpolarized quantum liquid specific to the multi-species
system, which exemplifies a link between the spin state and the inter-layer
electron correlation.  Even when the ground state is spin-polarized, the lowest
charge-excitation mode involves the spin when
the interlayer tunneling is considered.
\ \\


\newpage

Recently, much attention is focused on the double fractional quantum Hall (FQH)
system, in which two layers interact with each other as realized in double
quantum wells\cite{boebinger} or in wide single quantum wells\cite{suen}.
Specifically, Eisenstein {\em et al} have observed a FQH state at total Landau
level filling of $\nu = 1/2$ in a structure in which the inter-layer tunneling
is prohibited due to a barrier separating the two layers and yet the two layers
are coupled via Coulomb interactions.\cite{eisenstein92}
Usual practice in considering a double FQH system is to introduce a pseudospin
describing the layer degrees of freedom, while the real spin is neglected under
the assumption that they are fully polarized.  Then the spin-polarized double
FQH system mimicks the single FQH system of spin $1/2$ electrons.

However, already in single-layer FQH systems,\cite{qhe} the real spin degrees
of freedom dominate the electron correlation via Pauli's principle.
Namely, the ground state is spin fully polarized for odd-fraction Landau level
filling factor, while the ground state is spin unpolarized for some other
fractions as detected in
tilted-field experiments.
Thus the real question for the double FQH system is that how the inclusion of
the spin degrees of freedom dominate the physics, in which
the total spin should crucially affect the interplay of the intra-layer and
inter-layer electron correlations, which is controlled by the layer separation
($d/ \ell$) normalized by the magnetic length ($\ell = \sqrt{ c \hbar/ e B}$).

Motivated by this, we consider double-layer FQH systems of spin $1/2$
electrons.
 We shall show from a numerical study for finite systems that we do have
spin-unpolarized FQH state specific to the double-layer system of spin $1/2$
electrons.
Another motivation of the present Letter is to look into the low-lying
excitations in the coexistence of real and pseudo spins
in view of the recently emerged measurements of the excitations in the FQH
system from the inelastic light scattering.\cite{pinczuk92,pietilainen,longo}

 We consider the Hamiltonian, first in the absence of inter-layer tunneling,
given by
\begin{eqnarray}
 H = \frac 12 \sum _{m_1 \sim m_4} \sum _{\lambda _1 \sim \lambda _4}
 \sum _{\sigma, \sigma^{\prime}}
 \langle m_1 \lambda _1, m_2 \lambda _2 | \frac{e^2}{\epsilon \sqrt{| {\bf r}_1
- {\bf r}_2 |^2 + ( z_1 - z_2 )^2}} | m_4 \lambda _4, m_3 \lambda _3 \rangle
 a_{m_1 \lambda _1}^{\sigma \dagger} a_{m_2 \lambda _2}^{\sigma^{\prime}
\dagger} a_{m_3 \lambda _3}^{\sigma^{\prime}} a_{m_4 \lambda _4}^{\sigma},
\nonumber \\
\end{eqnarray}
 where $a_{m \lambda}^{\sigma \dagger}$ is the creation operator for the {\it
m}-th in-plane orbit of real spin $\sigma$ with pseudospin $\lambda (=$ layer
$1,2$),
${\bf r}$ the in-plane position and $\epsilon$ the dielectric constant.

We have obtained the ground-state wavefunctions from the exact diagonalization
of finite systems in both torus and spherical geometries.
 Since the total spin, $S_{\rm tot}$, of the system is conserved, we
concentrate on the subspace of $S_z^{\rm tot}= (N_{\uparrow} - N_{\downarrow})/
2 = 0$.
 Still, the inclusion of the spin in the double-layer system enormously
increases the dimension of the Hamiltonian matrix
(to typically $8\times 10^5$ for $\nu=4/7$ with four electrons per each layer),
which has been diagonalized here by the Lanczos method.
 We have determined $S_{\rm tot}$ and the intra- and inter-layer electron
correlations for various values of total $\nu$ and $d$.

To characterize the numerically obtained wavefunctions we have looked into, in
addition to the radial distribution function, the overlap with trial
wavefunctions: we can extend the Greek-Roman wavefunction\cite{halperin83}
proposed for the single-layer system of spin $1/2$ electrons to the
double-layer system, which is feasible
in the absence of inter-layer tunneling with the fixed number of electrons in
each layer.\cite{tomiura}
 The `double Greek-Roman' wavefunction is given (in the symmetric gauge) for
$N$ electrons as
\begin{eqnarray}
\Psi _{lmn} &=& \hat{\rm A}\ [\ \Phi _{lmn}(z)\ (u \alpha)_1 \cdots (u
\alpha)_{\frac N4}(u \beta)_1 \cdots (u \beta)_{\frac N4} \nonumber \\
& &\times \ (d \alpha)_1 \cdots (d \alpha)_{\frac N4}(d \beta)_1 \cdots (d
\beta)_{\frac N4} \ ], \\
\Phi _{lmn}(z)&=&\Phi_{\uparrow \uparrow}^{{\rm intra}} \Phi_{\uparrow
\downarrow}^{{\rm intra}} \Phi^{{\rm inter}} \exp{(-\sum _{i=1}^N|z_i|^2/
4\ell^2)}, \\
\Phi_{\uparrow \uparrow}^{{\rm intra}} &=&\prod _{1 \leq i < j \leq \frac N4}
(z_i - z_j)^l (Z_i - Z_j)^l (\xi_i - \xi_j)^l (\Xi_i - \Xi_j)^l, \\
\Phi_{\uparrow \downarrow}^{{\rm intra}} &=&\prod _{1 \leq i , j \leq \frac N4}
(z_i - Z_j)^m (\xi_i - \Xi_j)^m, \\
\Phi^{{\rm inter}} &=&\prod _{1 \leq i , j \leq \frac N4} (z_i - \xi_j)^n (z_i
- \Xi_j)^n (Z_i - \xi_j)^n (Z_i - \Xi_j)^n.
\end{eqnarray}
 Here $z_i = x_i-{\rm i}y_i$ is the position of an $\uparrow$-spin electron in
layer 1, $Z_i$ for a $\downarrow$-spin in layer 1, $\xi_i$ for a
$\uparrow$-spin in layer 2 and $\Xi_i$ for a $\downarrow$-spin in layer 2.
 $\hat{\rm A}$ is the antisymmetrization operator, $u/d$ are the spinors for
layer 1/2, $\alpha/\beta$ are the spinors for real spin up/down.
 The exponents in the Jastrow factors specify the orbital correlation (minimum
relative angular momentum) for intra-layer like spins $(l)$, intra-layer unlike
spins $(m)$ and inter-layer electrons $(n)$.

 Fermi statistics requires $l$ to be odd.  In addition a wavefunction must be
an eigenstate of $S_{\rm tot}$.
 Since the total spin of each layer is conserved in the absense of inter-layer
tunneling, we should impose the usual Fock condition upon each layer, which is
satisfied only when $m=l$ (spin-polarized) or $m=l-1$ (spin-unpolarized) for
$S_z^{\rm tot}=0$.
The filling factor is given by
$\nu = 4/(l + m + 2n),$
since we have $N_{\phi} = (l + m + 2n)\,N/ 4 - l$
in a spherical system with the number of flux quanta being $N_{\phi}$, which is
in turn related to the Landau-level filling via $N_{\phi} = \nu ^{-1}\,N -
\delta$ where $\delta$ is an integer.

We have previously obtained the result for total $\nu = 1$.\cite{tomiura}
The result shows an existence of a spin-polarized/spin-unpolarized transition
at $(d/\ell\,)_c = 1.43$.
A change in the inter-layer radial distribution function, which is
quantitatively slight but discontinuous, signals the transition at $d_c$.
The spin-polarized ground state for $\nu=1$ has a large overlap with $\Psi
_{111}$.  This is in fact expected,
 since we have an obvious limit of $d=0$ at which both the pseudospin and real
spin should be polarized with all the correlations between like/unlike layers
and like/unlike spins becoming equivalent as realized in $\Psi _{mmm}$.

Now we turn to the case where the role of spin is truly dramatic.  The double
Greek-Roman traial function predicts the simplest spin-unpolarized state to be
$\nu = 4/7 = 4/(3+2+2 \times 1)$, for which $\Psi _{321}$ becomes an eligible
function.
 This state is thus in sharp contrast with the extensively studied $\nu = 1/ 2$
state, in which case the only eligible function among $\Psi _{lmn}$'s with a
finite inter-layer correlation ($n\neq0$) is spin-polarized $\Psi _{331}$.
 While for $\nu = 1/ 2$ Yoshioka {\em et al} have pointed out, for spinless
electrons, a large overlap between the ground state and the (spinless
counterpart of) $\Psi _{331}$ around $d/\ell \simeq 1.5$ from the numerical
calculation \cite{ymg89}, followed by an experimental identification of the
state by
Eisenstein {\em et al},\cite{eisenstein92} the $\nu = 4/7$ state thus
exemplifies a novel, spin-unpolarized class in the coexistence of real and
pseudo-spins.

 As seen in Fig.1(a), preliminary reported in \cite{tomiura}, the overlap
between the exact ground state at  $\nu = 4/7$ and $\Psi _{321}$
numerically calculated in the spherical geometry has indeed a maximum value,
0.968, around $d/\ell \simeq 1.0$.
 $\Psi _{321}$ has the intra-layer correlation similar to that of the $\nu = 2/
5$ single-layer state with $S_{\rm tot} = 0$ proposed by
Halperin,\cite{halperin83} but incorporates a significant inter-layer
correlation as well.
Namely, given the fact that three of the intra-layer correlation of parallel
spins, intra-layer correlation of antiparallel spins and inter-layer
correlation cannot vary independently in a quantum liquid, $\Psi _{321}$
provides a simplest example in which all the three correlations differ from
each other.
 The state is realized for a finite range of $d$ because, as $d$ is increased,
the difference between the intra-layer and inter-layer interactions
(or the Haldane pseudopotentials for the spherical geometry) increases, thereby
giving a chance for the inter-layer correlation to deviate from the intra-layer
correlation, while the system will eventually reduce to independent layers for
larger $d$.

 Experimentally the quantization at $\nu = 4/7$ in double FQH systems has not
been observed so far.  This may be because the layer separation has not been
made small enough.
 Another factor is that a spin-unpolarized state will be unfavored when the
Zeeman energy, $E_{\rm Zeeman}$, is taken into account.
  Hence it is imperative to confirm the $\nu=4/7$ state can survive the Zeeman
effect.
 We have calculated the energy difference per particle, $\Delta E$, between the
lowest of the spin-polarized states and the ground state as a function of
$d/\ell$.
 The result in Fig.1(b)
shows that it has a peak of $0.0058 \,e^2/ \epsilon \ell$, which is $\simeq
1.0$K for GaAs (with $\epsilon = 12.6$) in $B=10$T and is in fact comparable
with the Zeeman energy $g \mu _B Bs \simeq 1.5$K, where $g (=0.44$ for GaAs) is
Land\'{e}'s $g$ factor.\cite{thick}
The figure may also serve as a phase diagram, if we normalize the vertical axis
to regard it as the ratio, $\Delta E/E_{\rm Zeeman}$
(the right scale in the figure): in the region where the curve exceeds unity
the spin-unpolarized ground state survives the Zeeman energy.
 Since $\Delta E/E_{\rm Zeeman}\propto (g\sqrt{B})^{-1}$ we predict that the
$\nu = 4/ 7$ FQH state should be observable for smaller $B$ (with smaller
density of electrons to retain $\nu = 4/ 7$), or for smaller $g$-factor
possibly realized in high-pressure experiments \cite{clarkp}.

We have also calculated the energy gap between the ground state and the first
excited state as a function of $d/\ell$.
 The result in Fig.1(a) has a peak of $0.027 e^2/ \epsilon \ell$, which
has a magnitude similar to the gap for the single-layer $\nu = 1/ 5$ state.
 All of the three curves in Fig.1
are peaked in the same region of $d$, which confirms the existence of an
intrinsic state in this region.

We can further show that, even when the ground state is real-spin polarized,
the discussion of charge excitations has to include the spin degrees of freedom
when we
take the interlayer tunneling into account.
The tunneling adds a term,
$H_t  =  - (\Delta _{\rm SAS}/2) \sum _{m, \sigma} ( a_{m 1}^{\sigma \dagger}
a_{m 2}^{\sigma} + {\rm h.c.} )$ ,
to the Hamiltonian.
The single-particle wavefunctions then split into symmetric and antisymmetric
ones about the center of the structure, and the gap, $\Delta_{\rm SAS}$, enters
as another energy.
For spin $1/2$ electrons in a double layer, we have then to consider the
excitation mode in which pseudospin-flip and real-spin-flip simultaneously take
place (which we call SPS-mode) in addition to
the spin-wave (S) and pseudospin-wave (PS) excitations.

  This is expected from the effective spin/pseudospin Hamiltonian for the
system, which comprises the pair creation/annihilation of the PS mode, the $S
\leftrightarrow PS + SPS$ and $SPS \leftrightarrow S + PS$ processes on top of
the free boson piece.
 Because of these processes, the effective Hamiltonian cannot be diagonalized
by a Bogoliubov transformation, unlike the spinless case where the PS mode may
be nicely fitted by the single-mode approximation.
Brey has discussed an SPS excitation in the Hartree-Fock
approximation.\cite{brey}
 However, this problem has to be investigated rigorously, because we are
dealing with a strongly correlated system.

Here we look into the case of $\nu=1$ because of the recent interests for this
situation in the presence of inter-layer tunneling.  Murphy {\em et al} \ have
experimentally investigated the double-layer system \cite{murphy}
 to probe the phase diagram, in which
the $\nu = 1$ `QHE region' exists with
the ground state being both real-spin polarized and pseudospin polarized
throughout.
There the nature of the ground state evolves continuously, as $\Delta _{\rm
SAS}$ is increased, from $\Psi_{111}$ dominated by the inter-layer Coulomb
interaction ((a) in the inset of Fig.2) down to the fully-occupied symmetric
state, $\Psi_{\rm sym}^{\nu=1}$, dominated by single-particle tunneling (b).

We present in Fig.2 the numerical result for the low-lying excitations for
three typical points in the QHE region, which includes the case (c) of a large
$d/\ell$ for which a dip evolves in the pseudospin-wave dispersion precursing
an instability of $\Psi_{\rm sym}^{\nu=1}$.
  When $\Delta _{\rm SAS}\neq 0$, the spin-wave excitation is a gapless
Goldstone mode restoring the SU(2) symmetry of the spin if the Zeeman shift is
neglected,
while both the PS mode and the SPS modes have gaps.  A new finding here is that
the energies of the three modes satisfy
 $E_S < E_{SPS} < E_{PS}$ for $k<1/\ell$ for cases (a-c), which persists when
the Zeeman energy is considered.
 Namely, the charge excitation costs smaller energy $(<\Delta _{\rm SAS})$ when
the spin excitation is exploited simultaneously.
 Thus the SPS-mode can be a candidate for the thermal gap, but this would
contradict with the experimentally reported gap much larger than $\Delta _{\rm
SAS}$ for samples having a small $\Delta _{\rm SAS}$ \cite{murphy}.
 The problem is thus still open.
 For the spin-polarized case, the pseudospin-wave excitations from the ground
state have been shown to have multiplet structures of weakly-interacting bosons
(pseudomagnons) for small layer separations.\cite{issp}
 Extension of this picture to spin 1/2 electrons in a double-layer is another
future problem.

 We are grateful to Prof. D. Yoshioka and Koichi Kusakabe for valuable
discussions.
 The numerical calculations were done on HITAC S3800 in the Computer Centre,
the University of Tokyo.
 This work was in part supported by a Grant-in-Aid from the Ministry of
Education, Science and Culture, Japan.
\par

\newpage
\noindent Figure captions\par
\ \\
\noindent {\bf Fig.1.}  (a)The overlap between the exact ground state at
$\nu=4/7$ and $\Psi _{321}$ against the layer separation, $d$
 for eight (four per layer) electrons in the spherical geometry.
  The energy gap per particle between the ground state (which is
spin-unpolarized) and the lowest spin-polarized state (b) or the gap between
the ground state and the first excited state (a, right scale) are also plotted.
The right scale in (b) represents the ratio, $(E_{\rm Smax}-E_0)/E_{\rm
Zeeman}$, where $\tilde{g}\equiv g/g_{\rm GaAs}$ and $\tilde{B}\equiv B/10$T:
in the region where the line exceeds unity the spin-unpolarized ground state
survives the Zeeman energy.
    The lines are guide to the eye.\par
\ \\
\noindent {\bf Fig.2.} The excitation modes for the double-layer system of spin
$1/2$ electrons having six electrons at $\nu = 1$ for (a) $d/ \ell = 0.5$,
$\Delta _{\rm SAS}$/($e^2/\epsilon \ell$)$ = 0.05$, (b) $d/ \ell = 0.5$,
$\Delta _{\rm SAS}$/($e^2/ \epsilon \ell$)$ = 0.20$ and (c) $d/ \ell = 1.5$,
$\Delta _{\rm SAS}$/($e^2/ \epsilon \ell$)$ = 0.20$ in the absence of the
Zeeman energy.
 The modes comprise the spin-wave excitation (S, dotted line), pseudospin-wave
excitation (PS, broken line), and pseudospin-wave excitation with one-spin-flip
(PSP, solid line).
 The lines are guide to the eye.
The inset indicates the positions of the three sets of parameters on a plot of
the numerically obtained overlap between the exact wavefunction and the
fully-occupied symmetric state against $d/ \ell$ and $\Delta _{\rm SAS}$.


\begin{thebibliography}{99}
\bibitem{thick} The energy difference is shown to become slightly smaller if we
take the finite thickness of the wavefunction normal to the layers into
consideration.
\end{thebibliography}
\end{document}